\documentclass[10pt,a4paper,twocolumn,english,prb,aps,showpacs,floatfix,groupedaddress,superscriptaddress]{revtex4-2}
\usepackage{graphicx,epsfig}
\usepackage{times}
\usepackage{graphics,dcolumn,bm,float}
\usepackage{amssymb,amsmath,rotate,color,dsfont}
\usepackage[breaklinks,colorlinks=true,urlcolor=blue,citecolor=blue,linkcolor=blue]{hyperref}
\usepackage{breakurl}

\begin{document}
\unitlength 1 cm
\newcommand{\be}{\begin{equation}}
\newcommand{\ee}{\end{equation}}
\newcommand{\bearr}{\begin{eqnarray}}
\newcommand{\eearr}{\end{eqnarray}}
\newcommand{\nn}{\nonumber}
\newcommand{\dagg}{{\dagger}}
\newcommand{\vpdag}{{\vphantom{\dagger}}}
\newcommand{\vecr}{\vec{r}}
\newcommand{\bs}{\boldsymbol}
\newcommand{\up}{\uparrow}
\newcommand{\down}{\downarrow}
\newcommand{\fns}{\footnotesize}
\newcommand{\ns}{\normalsize}
\newcommand{\cdag}{c^{\dagger}}

\definecolor{red}{rgb}{1.0,0.0,0.0}
\definecolor{green}{rgb}{0.0,1.0,0.0}
\definecolor{blue}{rgb}{0.0,0.0,1.0}

\title{Antiferromagnetic Chern insulator in centrosymmetric systems}

\author{Morad Ebrahimkhas}
\email{ebrahimkhas@itp.uni-frankfurt.de}
\affiliation{Goethe-Universit\"at Frankfurt, Institut f\"ur Theoretische Physik, Frankfurt, Germany}
\affiliation{Department of Physics, Mahabad Branch, Islamic Azad University, Mahabad, Iran}
\author{G\"otz S. Uhrig}
\email{goetz.uhrig@tu-dortmund.de}
\affiliation{Condensed Matter Theory, Department of Physics, TU Dortmund, 44221 Dortmund, Germany}
\author{Walter Hofstetter}
\email{hofstett@physik.uni-frankfurt.de}
\affiliation{Goethe-Universit\"at Frankfurt, Institut f\"ur Theoretische Physik, Frankfurt, Germany}
\author{Mohsen Hafez-Torbati}
\email{mohsen.hafez@tu-dortmund.de}
\affiliation{Condensed Matter Theory, Department of Physics, TU Dortmund, 44221 Dortmund, Germany}

\begin{abstract}
An antiferromagnetic Chern insulator (AFCI) can exist if the
effect of the time-reversal transformation on the electronic state cannot be compensated by a space
group operation. The AFCI state with collinear magnetic order is already realized in 
noncentrosymmetric honeycomb structures through the Kane-Mele-Hubbard model. 
In this paper, we demonstrate the existence of the collinear AFCI in a 
square lattice model which preserves the inversion symmetry. Our study relies on the time-reversal-invariant Harper-Hofstadter-Hubbard
model extended by a next-nearest-neighbor hopping term including spin-orbit coupling and a checkerboard potential.
We show that an easy $z$-axis AFCI appears between the band insulator at weak and the
easy $xy$-plane AF Mott insulator at strong Hubbard repulsion provided the checkerboard potential
is large enough.
The close similarity between our results and the results obtained for the noncentrosymmetric 
Kane-Mele-Hubbard model suggests the AFCI as a generic consequence of spin-orbit coupling 
and strong electronic correlation which exists beyond a specific model or lattice structure.
An AFCI with the electronic and the magnetic properties originating from the same strongly interacting
electrons is promising candidate for a strong magnetic blue shift of the charge gap 
below the N\'eel temperature and for realizing the quantum anomalous Hall effect at higher temperatures
so that applications for data processing become possible.

\end{abstract}

\maketitle

\section{Introduction}
The precise quantization of the Hall conductance in a two-dimensional (2D) electron 
gas system subject to a strong perpendicular magnetic field at low temperatures led to the 
discovery of the quantum Hall state.
A quantum Hall state is characterized by a non-zero topological invariant $\mathcal{C}$ known as Chern number. 
The state shows insulating behavior in the bulk and metallic behavior at the edges.
The gapless edge states are chiral, i.e., electrons at each edge can only propagate in a single direction,
either clockwise or anticlockwise. It is determined solely by the direction of the applied 
magnetic field. This prevents backscattering and permits dissipationless charge transport at the 
edges \cite{Hasan2010,Qi2011}. 

While a strong external magnetic field was essential in the discovery of the quantum Hall 
state, the Haldane model \cite{Haldane1988} provided a theoretical demonstration that a quantum Hall state can be achieved 
even without a net magnetic flux through the system. The only necessary condition is breaking of
the time-reversal symmetry. This suggested that the quantum Hall state 
can be an intrinsic feature of a material rather than an effect induced by an external magnetic 
field. Such a state is known as the quantum anomalous Hall state or Chern insulator (CI). 
From a practical point of view, the CI is highly favorable for establishing dissipationless
charge transport since it requires no strong external magnetic field \cite{Liu2016,Tokura2019}.

The extension of the Haldane model to a spinful time-reversal-invariant (TRI) model 
led to the prediction of the quantum spin Hall insulator (QSHI) \cite{Kane2005}.
In a QSHI \cite{Kane2005,Bernevig2006a} the Chern number for up and down spins is opposite due to the time-reversal symmetry and 
the total Chern number $\mathcal{C}=\mathcal{C}_\uparrow+\mathcal{C}_\downarrow$ vanishes. 
The QSHI is characterized by a $\mathbb{Z}_2$ topological invariant $\nu=(\mathcal{C}_\uparrow-\mathcal{C}_\downarrow)/2$ 
modulo 2. There are gapless edge states which have opposite chirality for opposite spins. 
The QSHI remains robust against spin-mixing Rashba-like spin-orbit couplings due to the Kramers degeneracy \cite{Hasan2010,Qi2011}. 
The theoretical prediction of the QSHI in quantum wells \cite{Bernevig2006b} was soon followed by an 
experimental verification \cite{Koenig2007}. 
The generalization of the QSHI to three dimensions 
led to the theoretical prediction and the experimental observation of 3D TRI
topological insulators. While the quantum Hall state is a result of strong external magnetic fields,
the TRI topological insulators stem from {strong} spin-orbit coupling \cite{Hasan2010,Qi2011}.

Despite the important role that the Haldane model played in the discovery of the TRI
topological insulators, the realization of the CI, for which the Haldane 
model was originally proposed, still remained elusive. This originated from the fact that the 
Haldane model was introduced more for theoretical aims rather than to be materialized.
We would like to point out that the Haldane model is simulated in optical lattices by 
creating artificial gauge fields for neutral atoms \cite{Jotzu2014}.

The realization of the CI in a crystalline material
requires two main ingredients: a strong spin-orbit coupling and a 
{spontaneous} breaking of the time-reversal symmetry via spontaneous magnetization \cite{Tokura2019,Liu2016}.
An effective strategy to reach these conditions is by doping a topological insulator 
with magnetic impurities. A topological insulator has the prerequisite of strong spin-orbit 
coupling and magnetic impurities introduce magnetism into the system. {A} CI {has been} 
realized by doping thin films of the topological insulator (Bi,Sb)$_2$Te$_3$ with the transition-metal 
element Cr \cite{Chang2013} following a theoretical 
proposal \cite{Yu2010}. A precise quantization of the Hall resistance is observed at the temperature $30$ mK at 
zero magnetic field \cite{Chang2013}. 
Improving the sample quality has allowed the realization of the CI at the higher temperature $300$ mK \cite{Kou2015}. 
The fact that the CI is realized at a temperature two orders of magnitude smaller than the Curie temperature of the material, $30$ K, is attributed 
to the strong inhomogenity induced into the system by magnetic doping \cite{Mogi2015}. The temperature is raised 
to $2$ K using a magnetic modulation doping technique \cite{Mogi2015}. 

There has been a large interest in recent years in finding topological insulators with intrinsic magnetic order \cite{Wang2021}
which would allow the elimination of the detrimental effect of disorder and an observation of the CI at higher temperature.
The efforts has 
led to the identification of several intrinsic magnetic topological insulators, among them, MnBi$_2$Te$_4$ 
has attracted the most attention 
\cite{Li2019a,Otrokov2019,Gong2019,Chen2019a,Wu2019,Hu2020,Hao2019,Chen2019,Li2019b,Swatek2020}. 
However, these systems {have} a very weak coupling between the magnetism and the electronic surface states. 
A magnetically-induced gap in the surface states which is essential to realize the CI at high temperatures is either missing or 
is extremely small \cite{Hao2019,Chen2019,Li2019b,Swatek2020}. The CI is observed on
thin films of MnBi$_2$Te$_4$ with an odd number of layers but still limited to the small temperature of $1.4$ K \cite{Deng2020}. 
This encourages further research to find the CI in {a} new class of systems.

The current realization of the CI is tied to ferromagnetic ordering \cite{Liu2016,Tokura2019}. 
One notes that although MnBi$_2$Te$_4$ is an antiferromagnet (A-type) the realization of the CI in 
its thin films \cite{Deng2020} is due to the top and the bottom ferromagnetic layers.
In addition, the magnetic and the electronic properties originate from electrons in different orbitals.
While strongly correlated electrons in a partially filled $3d$ or $4f$ orbital are responsible for the 
magnetic ordering, the non-interacting electrons in $6p$ and $5p$ orbitals mainly govern the electronic 
properties \cite{Li2019b}. The electronic bands are only indirectly affected by the magnetic ordering 
via an AF Kondo interaction \cite{Yoshida2016}. 

Antiferromagnets are a 
large class of materials with unique features. They create no stray field, they are robust 
against disturbing magnetic fields, and they {have} ultrafast spin dynamics in contrast to 
ferromagnets. These are the reasons behind the recent surge of interest for
developing a spintronics technology based on antiferromagnets \cite{Baltz2018}.
In addition, there {is} experimental and theoretical evidence of antiferromagnets displaying a strong coupling
between the magnetic and the electronic properties 
\cite{Diouri1985,FerrerRoca2000,Sangiovanni2006,Wang2009,Fratino2017,Bossini2020,Hafez-Torbati2021}. 
This is especially the case if the electrons defining the charge gap are the ones responsible for the magnetic ordering, i.e., 
strongly correlated \cite{Bossini2020,Hafez-Torbati2021}.
There is a noticable shift of the charge gap towards higher energies upon developing the AF order 
below the N\'eel temperature \cite{Bossini2020,Hafez-Torbati2021}. 
These features make identifying strongly correlated systems which can host
antiferromagnetic CI (AFCI) states of particular interest.
We emphasize that the AFCI discussed in this paper originates from the {\it spontaneous} time-reversal symmetry breaking
which is distinct from the AF quantum Hall insulator \cite{Ebrahimkhas2021,Vanhala2016}
in which the time-reversal symmetry is explicitly broken.

A ferromagnetic ordering makes a clear distinction between up and down spins. This allows 
one spin component to be in a trivial state with $\mathcal{C}_{\sigma}=0$ {and the other spin component} to be in 
a quantum Hall state with $\mathcal{C}_{\bar{\sigma}}\neq 0$, 
resulting in a CI with $\mathcal{C}=\mathcal{C}_{\uparrow}+\mathcal{C}_{\downarrow}\neq 0$. 
In contrast, such a distinction is not obvious in an antiferromagnet. This 
raises the question if there can exist an AFCI state. An AFCI can indeed exist if the effect
of the time-reversal transformation on the electronic state cannot be compensated by a space 
group operation. The absence of such an antiunitary symmetry is necessary for finding a
nonzero Chern number \cite{Mong2010,Jiang2018}.
The AFCI has already been identified in the Kane-Mele-Hubbard model 
which lacks inversion symmetry using a mean-field theory approximation \cite{Jiang2018}
and is also suggested for NiRuCl$_6$ based on the density-functional theory analysis \cite{Zhou2016}. 

In this paper, we consider a minimal TRI extension of the Harper-Hofstadter model {\cite{Harper1955,Hofstadter1976,Thouless1982}}
which allows us to realize both the QSHI and the trivial band insulator (BI) at half-filling.
In addition to the nearest-neighbor (NN) hopping $t$ the Hamiltonian involves a next-nearest-neighbor (NNN)
hopping $t'$ and a checkerboard potential $\Delta$. The combination of {time-reversal} symmetry
and {inversion} symmetry with respect to a lattice site leads to spin degenerate energy bands. 
We address the effect of the Hubbard interaction $U$ favoring a N\'eel AF Mott insulator 
on the system employing the dynamical mean-field theory (DMFT) method {\cite{Georges1996}}. 
We determine the phase diagram of the model in the $U$-$\Delta$ plane for a fixed value of $t'$ and 
show that for $(U\approx 2\Delta) \gg t$ an easy $z$-axis AFCI appears, separating the BI at small 
and the easy $xy$-plane AF Mott insulator at large values of $U$. 

The similarity between our results and the results obtained for the noncentrosymmetric Kane-Mele-Hubbard model \cite{Jiang2018} 
suggests the AFCI as a generic phase in strongly correlated systems with spin-orbit coupling, which exists beyond a specific 
model or lattice structure. 
We discuss the stabilization of the AFCI based on vanishing of spin and charge gaps and emphasize that the 
necessary condition to realize the AFCI is the absence of a space group operation to compensate the effect of the 
time-reversal transformation on the electronic state.

The paper is organized as follows. In the next section we introduce the extended TRI Harper-Hofstadter-Hubbard
(HHH) model and present its phase diagram. Section \ref{sec:ts} is devoted to the technical aspects.
In Section \ref{sec:sp} we discuss the results obtained for small to intermediate values of the checkerboard 
potential $\Delta$. We unfold the emergence of the AFCI for large values of $\Delta$ in Section \ref{sec:ci}.
Section \ref{sec:gap} discusses the evolution of the essential charge and spin gaps across the AFCI phase.
The paper is concluded in Section \ref{sec:con}.

\section{the Model and the phase diagram}
\label{sec:ehhh}

\subsection{Hamiltonian}
The Kane-Mele model \cite{Kane2005a} and the TRI Harper-Hofstadter model \cite{Goldman2010,Cocks2012} have been introduced as 
two fundamental models to study the QSHI. The TRI Harper-Hofstadter model is simulated using ultracold atoms in optical lattices \cite{Aidelsburger2013,Miyake2013}. 
While the Kane-Mele model at half-filling represents a QSHI, 
the Harper-Hofstadter model at half-filling is a semi-metal. {One needs to consider the Harper-Hofstadter model at} other fillings or to  
add additional terms such as a NNN hopping to open a gap and achieve 
non-trivial topological states \cite{Hatsugai1990}. We include a NNN hopping to the {TRI}
Harper-Hofstadter model and focus on {half-filling}. The half-filling is necessary to realize a Mott insulating phase in the 
presence of a strong Hubbard interaction and to make a comparison with the results of the Kane-Mele-Hubbard model \cite{Jiang2018}.

A minimal extension of the TRI HHH model which allows to investigate the competition between the QSHI, the 
BI, and the Mott insulator states is thus given by
\be
\label{eq:hhh}
H=H_t+\Delta \sum_{\vec{r},\sigma}(-1)^{x+y} n^\vpdag_{\vec{r},\sigma}
+U\sum_{\vec{r}} n^\vpdag_{\vec{r},\down} n^\vpdag_{\vec{r},\up}
\ee
with the hopping term
\begin{eqnarray}
\label{eq:ht}
   H_t=&-&t\sum_{\vec{r}} 
   \left( c^\dagger_{\vec{r}+\hat{x}} c^\vpdag_{\vec{r}} 
   +{c}^\dagger_{\vec{r}+\hat{y}} e^{+2\pi i \varphi x \boldsymbol{\sigma_{\!z}^\vpdag}} c^\vpdag_{\vec{r}} + {\rm H.c.} \right)
\nn \\
&-&t'\sum_{\vec{r}} 
\left( 
c^\dagger_{\vec{r}+\hat{x}+\hat{y}} 
e^{+2\pi i \varphi (x+\frac{1}{2})\boldsymbol{\sigma_{\!z}^\vpdag}}  c^\vpdag_{\vec{r}} \right. 
\nn \\ && \left.
\phantom{+++}
+ c^\dagger_{\vec{r}+\hat{y}} 
e^{+2\pi i \varphi (x+\frac{1}{2})\boldsymbol{\sigma_{\!z}^\vpdag}} c^\vpdag_{\vec{r}+\hat{x}} + {\rm H.c.}
\right) 
\end{eqnarray}
where $\vec{r}=x\hat{x}+y\hat{y}=(x,y)$ runs over the square lattice with the lattice constant considered as the unit of 
length. The creation field operator at the lattice position $\vec{r}$ is given by 
$c^\dagger_{\vec{r}}=( c^\dagger_{\vec{r},\uparrow}, c^\dagger_{\vec{r},\downarrow} )$.
Similarly, the annihilation field operator is given by the column vector 
$c_{\vec{r}}=\left( c_{\vec{r},\uparrow}, c_{\vec{r},\downarrow} \right)^{T}$. 
The index $\sigma=\uparrow,\downarrow$ specifies the $z$-component of the particle spin. 
The occupation number operator reads 
$n^\vpdag_{\vec{r},\sigma}=c^\dagger_{\vec{r},\sigma} c^\vpdag_{\vec{r},\sigma}$. 

\begin{figure}[t]
   \begin{center}
   \includegraphics[width=0.32\textwidth,angle=0]{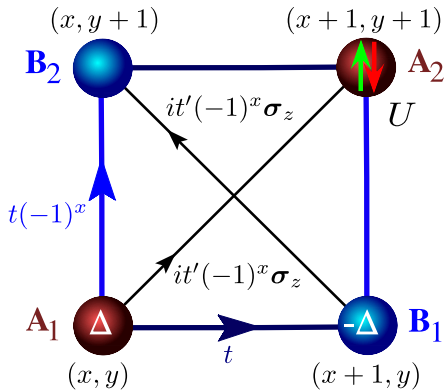}
   \caption{Schematic representation of the Hamiltonian \eqref{eq:hhh} for the phase parameter $\varphi=1/2$. 
   The Hamiltonian involves a nearest-neighbor hopping $t$, a next-nearest-neighbor hopping $t'$, a 
   checkerboard potential $\Delta$, and a Hubbard interaction $U$. {Particles} with opposite spins pick up
   opposite phases upon hopping due to the Pauli matrix $\boldsymbol{\sigma_{\!z}^\vpdag}$ in the phase factor. The different sublattices
   are labeled as $A_1$, $A_2$, $B_1$, and $B_2$.}
   \label{fig:hhh}
   \end{center}
\end{figure}

A particle picks up a phase determined by $\varphi$ upon hopping which is opposite for opposite spins 
due to the Pauli matrix $\boldsymbol{\sigma_{\!z}^\vpdag}$, making the Hamiltonian time-reversal symmetric. 
We fix the phase parameter to $\varphi=1/2$ and the NNN hopping to $t'=0.25t$ throughout this paper. 
The phase parameter $\varphi=1/2$ is the simplest case to achieve nontrivial topological bands.
The second term in Eq.\ \eqref{eq:hhh} is a checkerboard potential giving the onsite energy $+\Delta$
to  {sites} with $x+y$ even and the onsite energy $-\Delta$ to {sites} with $x+y$ odd.
It supports a BI phase in the system. 
The last term in Eq.\ \eqref{eq:hhh} is the Hubbard interaction which describes the repulsion between particles 
with opposite spins occupying the same lattice site and {at half filling} favors long-range AF order. 
The different terms in the Hamiltonian \eqref{eq:hhh} 
for $\varphi=1/2$ are sketched schematically in the $2\times2$ unit cell in Fig.\ \ref{fig:hhh}.
We denote the four different sublattices as $A_1$ with $x$ and $y$ even, $A_2$ with $x$ and $y$ odd, 
$B_1$ with $x$ odd and $y$ even, and $B_2$ with $x$ even and $y$ odd
\footnote{It is possible to reduce the number of sites in the unit cell to two sites by applying
the spin-dependent local gauge transformation 
$c_{\vec{r}} \rightarrow e^{-i\pi \boldsymbol{\sigma_{\!z}^\vpdag} y^2 /2} c_{\vec{r}}$. 
However, this shifts the energy bands in the momentum space and also changes the magnetic order in 
the large Hubbard $U$ limit. We decided to avoid such a transformation to prevent potential confusion.}. 
The system has inversion symmetry where the center of inversion can be at any lattice site.

A clarification on the choice of the staggered potential in Eq.\ \eqref{eq:hhh} is in order. We have 
considered a checkerboard potential changing as $\Delta(-1)^{x+y}$ and not a staggered potential changing 
only along one direction, e.g., $\Delta(-1)^{x}$.
This is because in the presence of the checkerboard potential the 
effect of the time-reversal transformation on the N\'eel AF state cannot be compensated 
by a space group operation while for the staggered potential along the $x$ direction the
effect of the time-reversal transformation on the N\'eel AF state can be compensated 
by a lattice shift along the $y$ direction.
The checkerboard potential permits the emergence of the N\'eel AFCI. The staggered potential changing only along the $x$ direction 
can lead to other types of correlated topological states and we leave it for a future study. The Hamiltonian \eqref{eq:hhh}
with $\varphi=1/2$ is thus motivated as a minimal theoretical model which allows to investigate the AFCI phase beyond the 
noncentrosymmetric Kane-Mele-Hubbard model \cite{Jiang2018}.

\subsection{Limiting behaviors}
In the absence of the Hubbard interaction the Hamiltonian \eqref{eq:hhh} reduces to a four-level problem in momentum
space described by
\be
\label{eq:hu0}
H=\sum_{\vec{k},\sigma} \Phi_{\vec{k},\sigma}^{\dagger} \boldsymbol{\mathcal{H}}_{\sigma}(\vec{k}) \Phi_{\vec{k},\sigma}^{\vpdag}
\ee
with $\Phi_{\vec{k},\sigma}
=(a_{1,\vec{k},\sigma},b_{1,\vec{k},\sigma},
b_{2,\vec{k},\sigma},a_{2,\vec{k},\sigma})^T$ where $a_{1,\vec{k},\sigma}$ is the Fourier transform of 
$c_{\vec{r},\sigma}$ on the sublattice $A_1$ and similarly for the other operators. The Bloch Hamiltonian is
given by
\bearr
\label{eq:hk}
\boldsymbol{\mathcal{H}}_{\sigma}(\vec{k})&=&\left(\Delta \boldsymbol{\sigma_{\! z}^\vpdag}
-2t\cos(k_y)\boldsymbol{\sigma_{\! x}^\vpdag}\right)\otimes \boldsymbol{\sigma_{\!z}^\vpdag}
-2t\cos(k_x)\boldsymbol{\mathds{1}}\otimes \boldsymbol{\sigma_{\!x}^\vpdag}  \nn \\
&&+4t'{\rm sgn}(\sigma)\sin(k_x)\sin(k_y)\boldsymbol{\sigma_{\!x}^\vpdag}\otimes \boldsymbol{\sigma_{\!y}^\vpdag}
\eearr
where $\boldsymbol{\sigma_{\!x}^\vpdag}$, $\boldsymbol{\sigma_{\!y}^\vpdag}$, and 
$\boldsymbol{\sigma_{\!z}^\vpdag}$ stand for Pauli matrices and we have defined ${\rm sgn}(\uparrow)=+1$ and 
${\rm sgn}(\downarrow)=-1$. 

The energy bands of Eq.\ \eqref{eq:hk} are spin degenerate as $\boldsymbol{\mathcal{H}}_{\uparrow}(\vec{k})$ and 
$\boldsymbol{\mathcal{H}}_{\downarrow}(\vec{k})$ differ only by a transpose. 
This spin degeneracy originates from the time-reversal symmetry, 
$\boldsymbol{\mathcal{H}}_{\sigma}(\vec{k})=\boldsymbol{\mathcal{H}}^*_{\bar{\sigma}}(-\vec{k})$, and the inversion symmetry 
with respect to a lattice site, 
$\boldsymbol{\mathcal{H}}_{\sigma}(\vec{k})=\boldsymbol{\mathcal{H}}_{\sigma}(-\vec{k})$. 
We have used $\bar{\sigma}$ to indicate the opposite direction of $\sigma$.

The Hamiltonian \eqref{eq:hk} at half filling describes a QSHI for $\Delta<4t'$ with the Chern numbers 
$\mathcal{C}_{\uparrow}=+1$ and $\mathcal{C}_{\downarrow}=-1$ for the up and down spins. Upon increasing 
$\Delta$ the gap closes at $\Delta=4t'$ and for $\Delta>4t'$ the system becomes a BI with $\mathcal{C}_\sigma=0$
for both spin components. As we will show using the topological Hamiltonian method \cite{Wang2012,Wang2013} 
a similar Hamiltonian but with a renormalized staggered potential determines the topological properties 
also of the interacting model \eqref{eq:hhh}.

In the large Hubbard $U$ limit {where} the subspace with a finite number of {doubly occupied sites} is much higher in energy 
than the subspace with no double occupancy one can derive an effective spin Hamiltonian describing 
the low-energy properties of the system \cite{Takahashi1977}. 
The effective spin Hamiltonian of the extended TRI HHH model \eqref{eq:hhh} in the Mott limit $(U-2\Delta)\gg t$ 
is given by
\be
\label{eq:hs}
H_{\rm eff}=J_1\sum_{\langle i,j \rangle} \vec{S}_i \cdot \vec{S}_j
+J_2\sum_{\left[ i,j \right]} \left( S^z_i S^z_j-S^x_i S^x_j -S^y_i S^y_j  \right)
\ee
with the NN $J_1=4t^2U/(U^2-4\Delta^2)$ \cite{Nagaosa1986a} and the NNN $J_2=4{t'}^2/U$ \cite{Rachel2010,Losada2019} 
exchange couplings. 
The notation $\langle i,j \rangle$ limits sites $i$ and $j$ to be NN and the notation $\left[ i,j \right]$ limits sites $i$ and $j$ to be NNN.
The spin anisotropy in the NNN exchange interaction stems from the spin-dependent NNN hopping
which reduces the ${\rm SU}(2)$ symmetry to the ${\rm U}(1)$ symmetry.
The Hamiltonian \eqref{eq:hs} supports the N\'eel AF state with a spin polarization in the $x$-$y$ plane ($xy$-AF state)
due to the ferromagnetic NNN exchange interaction in the $x$ and $y$ directions. This avoids 
the frustration induced by the AF NNN interaction in the $z$ direction. The energy difference
per lattice site of the N\'eel AF state with the spin polarization along the $z$ direction ($z$-AF state) and the $xy$-AF state
reads
\be
\label{eq:e0}
\varepsilon_z-\varepsilon_{xy}=J_2=\frac{4t'^2}{U}
\ee
in the mean-field approximation of the spin Hamiltonian \eqref{eq:hs}. While in the Mott regime $(U-2\Delta)\gg t$ 
the $xy$-AF state is the stable phase, we will show how charge fluctuations for $(U \sim 2\Delta)\gg t$ 
stabilize a $z$-AFCI. 

A similar Hamiltonian as Eq.\ \eqref{eq:hhh} but with hopping phases independent of spin, breaking explicitly the time-reversal symmetry, 
is employed in Ref.\ \cite{Ebrahimkhas2021} to investigate the role of the lattice symmetry on the 
emergence of AF quantum Hall states. One notes that here we analyze systems which show spontaneous breaking of the 
time-reversal symmetry and the emergence of AFCIs.
It is worth also mentioning that in the absence 
of the hopping phase, i.e. for $\varphi=0$, and vanishing NNN hopping $t'=0$, the Hamiltonian in Eq.\ \eqref{eq:hhh}
reduces to the ionic Hubbard model which has been used as a {paradigmatic} model for studying the transition from
band to Mott insulator in one \cite{Nagaosa1986a,Fabrizio1999,Hafez2014,Hafez2015,Loida2017} and 
two \cite{Garg2006,Ebrahimkhas2011,Kancharla2007,Paris2007,Byczuk2009,Hafez-Torbati2016} dimensions.

\begin{figure}[t]
   \begin{center}
   \includegraphics[width=0.47\textwidth,angle=0]{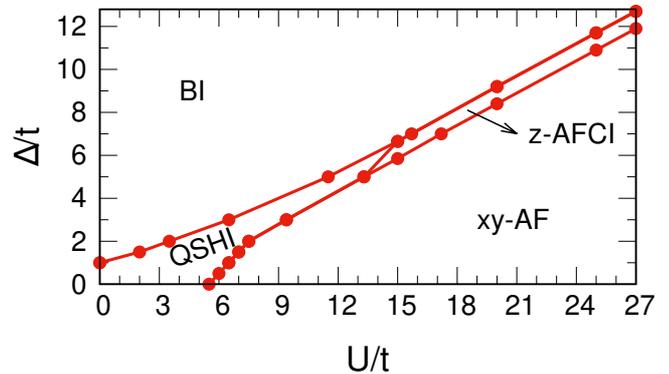}
   \caption{The phase diagram of the extended time-reversal-invariant Harper-Hofstadter-Hubbard model in Eq.\ \eqref{eq:hhh} 
   for the phase parameter $\varphi=1/2$ and the next-nearest-neighbor hopping $t'=0.25t$. The phase diagram involves 
   a trivial band insulator (BI), a quantum spin-Hall insulator (QSHI), a N\'eel antiferromagnetic 
   state with the spin polarization in the $xy$-plane ($xy$-AF), and a N\'eel antiferromagnetic Chern insulator
   with the spin polarization along the $z$-direction ($z$-AFCI). The results are for the temperature $T=0.02t$.}
   \label{fig:pd}
   \end{center}
\end{figure}

\subsection{Phase diagram}
Before proceeding to the technical aspects and discussing the results in details we present the 
phase diagram of the extended TRI HHH model \eqref{eq:hhh} for the phase parameter $\varphi=1/2$ and the NNN 
hopping $t'=0.25t$ in the $U$-$\Delta$ plane in Fig.\ \ref{fig:pd}. 
We have used {DMFT} \cite{Georges1996} and the topological 
Hamiltonian method \cite{Wang2012,Wang2013} to address the effect of the interaction. 
The system is {a} QSHI for small values of $\Delta$ and the Hubbard interaction induces 
a transition to the $xy$-AF phase. 
For intermediate values of $\Delta$ the system is in the BI phase and the Hubbard interaction first drives the  
system into the QSHI and then into the $xy$-AF state. For large values of $\Delta$, a $z$-AFCI separates 
the BI from the $xy$-AF phase. We cannot exclude the possible existence of an extremely narrow 
trivial $z$-AF phase between the BI and the $z$-AFCI.
The positions of the tri-critical points in Fig.\ \ref{fig:pd} are approximate as their accurate determination is demanding.
The phase diagram \ref{fig:pd} is obtained for the small but finite temperature
$T=0.02t$ to {guarantee} that the narrow $z$-AFCI phase can survive thermal fluctuations and be found in a 
real experiment which is always performed at a finite $T$.

The similarity between the phase diagram in Fig.\ \ref{fig:pd} and the phase diagram 
of the noncentrosymmetric Kane-Mele-Hubbard model \cite{Jiang2018} suggests that the $z$-AFCI is not 
a phase limited to a specific model or a specific lattice structure, but rather is a generic consequence 
of strong electronic correlation and spin-orbit coupling in two-dimensional systems. 
The necessary condition to realize an AFCI is 
the absence of a space group operation to compensate the effect of the time-reversal transformation on
the electronic state which can be achieved with or without the inversion symmetry. 
Our DMFT analysis demonstrates that 
the $z$-AFCI can exist beyond the (slave boson) mean-field approximation \cite{Jiang2018} and survives 
local quantum fluctuations.

\section{technical aspects}
\label{sec:ts}

\subsection{Dynamical mean-field theory}
We employ {DMFT} which is an established approach to systems with strong local interaction and large 
coordination number. The method approximates the self-energy to be spatially local which is exact in the limit 
of infinite coordination number. In the case of finite coordination number the non-local quantum 
fluctuations due to the momentum dependence of the self-energy are neglected. 
The local quantum fluctuations are fully taken into account.
The lattice model is mapped to an Anderson impurity model determined self-consistently \cite{Georges1996}.
{DMFT has been} extensively applied in the past decade to interacting two-dimensional topological systems
\cite{Cocks2012,
Budich2013,Amaricci2015,Irsigler2019,Hafez-Torbati2020,Ebrahimkhas2021,Vanhala2016,
Irsigler2019a}. 

We use {real-space} DMFT \cite{Potthoff1999,Song2008,Snoek2008} as it allows access not only to 
the bulk but also to the edges of the system. The self-energy is local but it can depend on the position,
\be
\label{eq:se}
\Sigma_{\vec{r},\sigma;\vec{r}',\sigma'}^\vpdag (i\omega_n)=
\delta_{\vec{r},\vec{r}'} \Sigma_{\vec{r};\sigma,\sigma'}^\vpdag(i\omega_n) \ .
\ee
One notes that the {spin} off-diagonal elements of the self-energy, 
$\Sigma_{\vec{r};\sigma,\bar{\sigma}}^\vpdag (i\omega_n)\neq 0$, are essential to describe the 
$xy$-AF phase. We use our implementation of {real-space} DMFT for SU($N$) systems introduced 
in Ref.\ \cite{Hafez-Torbati2018}
due to its easy adaptability to different models and its capability of addressing 
self-energies with spin off-diagonal elements. We use the exact diagonalization (ED) method as the 
impurity solver which provides an accurate description of static local quantities and permits 
direct access to real-frequency dynamical quantities \cite{Caffarel1994}. 

We mainly consider $40\times40$ lattices with periodic boundary conditions to investigate the bulk 
properties. However, near phase transitions especially in the $z$-AFCI region we have performed calculations 
also for $60\times60$ lattices and find no change in the results. 
To analyze edge excitations, $41\times40$ lattices are used with open boundary conditions along $x$ 
and periodic boundary conditions along $y$ direction, i.e., cylindrical geometry.  
One notes that although we use the real-space DMFT we fully exploit the translational symmetry 
of the system to set up the impurity problem and to compute the lattice Green's function \cite{Hafez-Torbati2018}.
For example, for the bulk properties the impurity model is addressed only at the four lattice 
sites in the unit cell sketched in Fig.\ \ref{fig:hhh}.

We mainly use the number of bath sites $n_{bs}=6$ in the effective Anderson impurity problem. However, we will present 
data also for $n_{bs}=5$, $7$, and $8$ across the $z$-AFCI phase and show that the results obtained 
for these different numbers of bath sites are almost indistinguishable. This stems from the perfect 
description of the dynamical Weiss field we find in all {these} cases.
The conservation of the total $z$-component of spin is used in the ED except for the $xy$-AF phase.
We fix the temperature to $T=0.02t$ and perform the DMFT loop with 200 positive Matsubara frequencies.
Again, we pay careful attention to the $z$-AFCI region {where we increase} the number of positive Matsubara 
frequencies up to 1000. A chemical potential $\mu$ is added to the Hamiltonian \eqref{eq:hhh} and is adjusted 
within the DMFT loop to satisfy the half-filling condition.

\subsection{Topological Hamiltonian method}
Although the real-space DMFT makes it possible to directly spot the excitations at the bulk and at the edges, 
a more precise way to determine the topological phase transitions is by computing the topological invariants. 
For an interacting system the topological invariants can be calculated using 
twisted boundary conditions \cite{Niu1985}. This requires knowledge of the eigenstates of the system which is 
beyond the scope of the DMFT. Topological invariants can be expressed also based on the Green's function 
provided the system contains no non-trivial degeneracy, such as the one in fractional quantum Hall 
states \cite{Ishikawa1986,Wang2010}.
Such a condition is satisfied for all the phases we investigate in this paper. 
The method involves frequency and momentum integrations over the Green's function and its derivatives and remains 
{computationally} demanding. In Ref.\ \cite{Wang2012} it {has been} shown by adiabatic deformation of the imaginary-frequency Green's function
(such that the charge gap never closes) that the topological invariant of an interacting system can be determined from 
an effective non-interacting model. The method is exact provided that the necessary conditions are satisfied \cite{Wang2013}. 
The effective model, called topological Hamiltonian, in the Bloch form reads
\be
\label{eq:th}
\boldsymbol{\mathcal{H}}_{\rm t}(\vec{k})=\boldsymbol{\mathcal{H}}_0(\vec{k})+\boldsymbol{\Sigma}(\vec{k},\omega=0),
\ee
where $\boldsymbol{\mathcal{H}}_0(\vec{k})$ is the non-interacting part of the Hamiltonian and 
$\boldsymbol{\Sigma}(\vec{k},0)$ is the self-energy in momentum space and at zero frequency.  

It should be mentioned that the topological invariant  
can change not only due to the poles of the Green's function but also due to
its zeros \cite{Gurarie2011}, e.g., in paramagnetic Mott insulators where the self-energy at zero 
frequency diverges \cite{Yoshida2016}. But this does not occur in the phases studied in this paper.

The non-interacting part of the Hamiltonian \eqref{eq:hhh} in the Bloch representation is given by Eq.\ \eqref{eq:hk}.
The self-energy in the DMFT is local. Hence, the momentum-space self-energy in the paramagnetic and in the $z$-AF state 
is given by 
\bearr
\label{eq:sek}
\boldsymbol{\Sigma}^\vpdag_{\sigma}(\vec{k},0)=
&+&\frac{1}{2} \left( 
\Sigma^\vpdag_{A,\sigma}(0)+\Sigma^\vpdag_{B,\sigma}(0)
\right) \mathds{1}\otimes\mathds{1} \nn \\
&+& \frac{1}{2} \left( 
\Sigma^\vpdag_{A,\sigma}(0)-\Sigma^\vpdag_{B,\sigma}(0)
\right) 
\boldsymbol{\sigma_{\! z}^\vpdag}\otimes\boldsymbol{\sigma_{\! z}^\vpdag} \ ,
\eearr
where $\Sigma_{A,\sigma}(0)$ is the self-energy on the sublattice with the higher onsite energy and  
$\Sigma_{B,\sigma}(0)$ is the self-energy on the sublattice with the lower onsite energy. One notes 
that the self-energy on the sublattices $A_1$ and $A_2$ are equal, and also on the sublattices $B_1$ and $B_2$.

The topological Hamiltonian obtained by adding the self-energy \eqref{eq:sek} to the Bloch Hamiltonian \eqref{eq:hk} 
remains, up to an irrelevant constant, the same as Eq.\ \eqref{eq:hk} but with the renormalized staggered potential
\be
\label{eq:rp}
\tilde{\Delta}_\sigma^\vpdag =\Delta+
\frac{1}{2} \left( 
\Sigma^\vpdag_{A,\sigma}(0)-\Sigma^\vpdag_{B,\sigma}(0) 
\right) \ .
\ee
We drop the lower index $\sigma$ when discussing paramagnetic phases 
as there is no spin dependence. 
The system is in the QSHI phase for $|\tilde{\Delta}|<4t'$ and in the BI phase 
for $|\tilde{\Delta}|>4t'$. In the $z$-AF state, however, the renormalized staggered potential \eqref{eq:rp} depends on 
the spin and, in principle, it is possible that one spin component falls in the topological region 
$|\tilde{\Delta}_{\sigma}|<4t'$ and the other in the trivial region $|\tilde{\Delta}_{\sigma}|>4t'$.
This leads to the emergence of a CI. For the NNN hopping $t'=0.25t$ that we consider throughout this paper the 
topological transition occurs at $|\tilde{\Delta}_{\sigma}|=t$.

The renormalized staggered potential in Eq.\ \eqref{eq:rp} requires the self-energy at zero frequency.
We always find that the value of the real-part of the self-energy at the smallest Matsubara frequency $\omega_0=\pi T$
perfectly describes the value of the self-energy at zero frequency obtained by a polynomial extrapolation. This means
one can replace $\Sigma_{A,\sigma}(0)$ with ${\rm Re}[\Sigma_{A,\sigma}(\omega_0)]$ 
and $\Sigma_{B,\sigma}(0)$ with ${\rm Re}[\Sigma_{B,\sigma}(\omega_0)]$ in Eq.\ \eqref{eq:rp}.

The DMFT together with the topological Hamiltonian method {have been} used extensively to map out 
the phase diagram of {various} interacting topological systems 
\cite{Budich2013,Amaricci2015,Irsigler2019,Hafez-Torbati2020,Ebrahimkhas2021,Vanhala2016,Amaricci2016}.
The phase diagram of the Haldane-Hubbard model obtained using 
the DMFT and the topological Hamiltonian method agrees qualitatively with the results obtained 
using ED and twisted boundary conditions \cite{Vanhala2016}. 
A systematic study of the quantum fluctuations beyond the DMFT indicates small changes to 
the phase boundaries \cite{Mertz2019}. Similarly, we expect our phase diagram to be qualitatively
reliable, i.e., to be reliable with regard to the types of the phases present and to which phase 
is adjacent to which other phase. But the position of 
the phase boundaries might be shifted upon including the nonlocal quantum fluctuations.

\section{weak to intermediate staggered potentials}
\label{sec:sp}

\begin{figure}[t]
   \begin{center}
   \includegraphics[width=0.47\textwidth,angle=0]{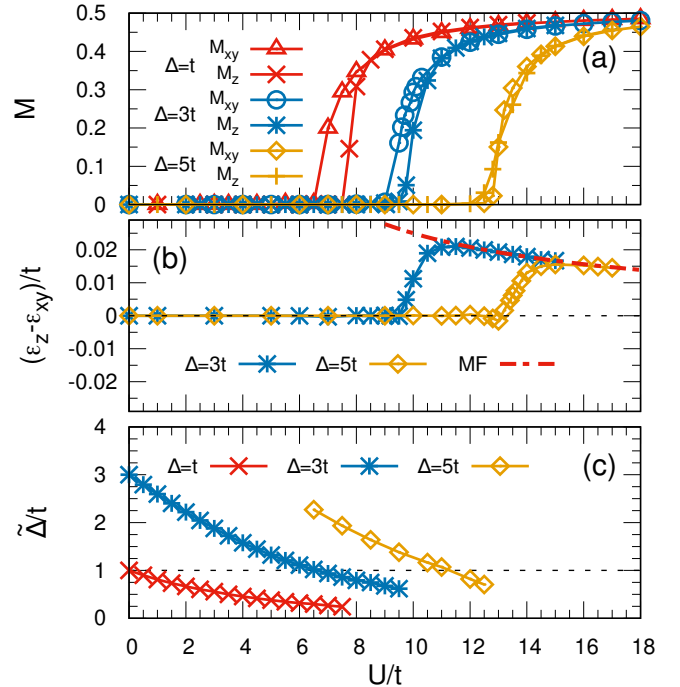}
   \caption{(a) The local spin polarization in the $xy$-AF state ($M_{xy}$) and in the $z$-AF state ($M_z$)
   vs the Hubbard interaction $U$ for different values of the staggered potential $\Delta$. (b) The energy 
   difference per lattice site of the $z$-AF state and the $xy$-AF state vs the Hubbard $U$ for $\Delta=3t$ 
   and $5t$. The mean-field {result} \eqref{eq:e0} of the effective low-energy spin model is included as a
   red dashed line for comparison. 
   (c) The renormalized staggered potential $\tilde{\Delta}$ in the paramagnetic region 
   vs the Hubbard $U$ for different values of $\Delta$. The system is in the trivial band insulator for 
   $\tilde{\Delta}>t$ and in the quantum spin-Hall insulator for $\tilde{\Delta}<t$ {which are separated by the dashed line at 
   $\tilde{\Delta}=t$}.
   The results shown are for the number of bath sites $n_{bs}=6$.}
   \label{fig:wd}
   \end{center}
\end{figure}

In Fig.\ \ref{fig:wd}(a) the local spin polarization in the $z$-AF state ($M_z$) and in
the $xy$-AF state ($M_{xy}$) is plotted vs the Hubbard interaction $U$ for the staggered potential 
$\Delta=t$, $3t$, and $5t$, which correspond to small and intermediate values of $\Delta$ in 
the phase diagram \ref{fig:pd}.
The results are for the number of bath sites $n_{bs}=6$ in the ED impurity solver. 
The local spin polarizations are given by
\begin{subequations}
\begin{align}
 M_z&=\frac{1}{2}|\langle c_{\vec{r}} ^\dagger ~ \boldsymbol{\sigma_{\! z}^\vpdag} ~c^\vpdag_{\vec{r}}  \rangle | \ , \\
 M_{xy}&=\frac{1}{2}|\langle 
 c^\dagger_{\vec{r}}~\boldsymbol{\sigma_{\! x}^\vpdag}~ c^\vpdag_{\vec{r}} \rangle \hat{x}
+ \langle c^\dagger_{\vec{r}}~ \boldsymbol{\sigma_{\! y}^\vpdag} ~c^\vpdag_{\vec{r}} \rangle \hat{y} | \ .
\end{align}
\end{subequations}
The lattice has the N\'eel AF order. The $xy$-AF phase is continuously degenerate
due to the spontaneous breaking of the ${\rm U}(1)$ symmetry. We have produced multiple $xy$-AF
solutions for the DMFT equations corresponding to the local spin polarization vector pointing 
in different directions in the $x$-$y$ plane. We find all the solutions having the same local spin 
polarization $M_{xy}$ and the same energy. This serves as a corroborating test for our results.
The $z$-AF solution is two-fold degenerate due to the spontaneous breaking of the 
time-reversal symmetry.

The $xy$-AF state and the $z$-AF state in Fig.\ \ref{fig:wd}(a) {show} the same local spin polarization 
at large values of $U$. 
However, the spin polarizations $M_{xy}$ and $M_{z}$ become distinct as the Hubbard $U$ is reduced. 
For small values of the staggered potential $\Delta$ the $xy$-AF state persists 
to smaller values of $U$ {compared} to the $z$-AF state. As one can see for $\Delta=t$ in Fig.\ \ref{fig:wd}(a) 
the spin polarization in the $z$-AF state vanishes at $U\sim 7.5t$ while the spin polarization in the $xy$-AF 
state survive down to $U\sim 6.5t$. This {enhanced} robustness of the $xy$-AF phase over the $z$-AF phase against 
the charge fluctuations reduces upon increasing the staggered potential to $\Delta=3t$ and disappears for $\Delta=5t$.

We have compared the energy per site of the $z$-AF ($\varepsilon_z$) and the $xy$-AF ($\varepsilon_{xy}$) states 
in Fig.\ \ref{fig:wd}(b) for $\Delta=3t$ and $5t$. One can see that the $xy$-AF state has a lower 
energy {compared} to the $z$-AF state. 
This is what one would naturally expect in the Mott regime based on the effective spin model \eqref{eq:hs}.
We expect our DMFT treatment in the Mott limit to be equivalent to the mean-field approximation of the low-energy 
spin model \eqref{eq:hs}. This is because the local quantum fluctuations taken into account in the DMFT influence 
only the high-energy properties, which are already eliminated in the derivation of the spin model.
The spin model only involves non-local quantum fluctuations. Such an equivalence can also be seen for the 
N\'eel temperature of the Hubbard model \cite{Rohringer2018}. The energy difference 
in Fig.\ \ref{fig:wd}(b) for large Hubbard interactions approaches  
the mean-field result of the spin model given by Eq.\ \eqref{eq:e0} with $t'=0.25t$, independent from $\Delta$.
The mean-field results are denoted in Fig.\ \ref{fig:wd}(b) by a dashed red line. 
One notes that although in the Mott regime $(U-2\Delta)\gg t$ the DMFT results equal the mean-field results of 
the effective low-energy spin model, the DMFT analysis takes into account the charge fluctuations which become essential 
as the Hubbard interaction is reduced.
We will show how these charge fluctuations for $(U \sim 2\Delta)\gg t$ favor energetically the $z$-AF state 
over $xy$-AF state. Evidence of the $z$-AF phase acquiring an energy lower than the $xy$-AF phase 
can already be seen in Fig.\ \ref{fig:wd}(b) for $\Delta=5t$ near $U=13t$. This becomes more pronounced
as $\Delta$ is further increased in Section \ref{sec:ci}. One notes that the location of the 
tri-critical points in the phase diagram \ref{fig:pd} are only approximate as they are difficult to be
determined accurately. 
 
In order to identify the transition between the BI and the QSHI in the paramagnetic region 
the renormalized staggered potential \eqref{eq:rp} is plotted in Fig.\ \ref{fig:wd}(c) vs the Hubbard $U$ for 
the different values of the staggered potential $\Delta=t$, $3t$, and $5t$. For $\Delta=t$ the system
is at the phase boundary of the BI and the QSHI for $U=0$. As the Hubbard $U$ is {increased} the renormalized 
staggered potential $\tilde{\Delta}$ lowers and the system becomes a QSHI. For $\Delta=3t$ and 
$\Delta=5t$ the system for small values of $U$ is in the BI phase with $\tilde{\Delta}>t$. 
As the Hubbard $U$ is increased the renormalized staggered potential crosses the dashed line at $\tilde{\Delta}=t$ 
and the transition to the QSHI takes place.

\begin{figure}[t]
   \begin{center}
   \includegraphics[width=0.434\textwidth,angle=0]{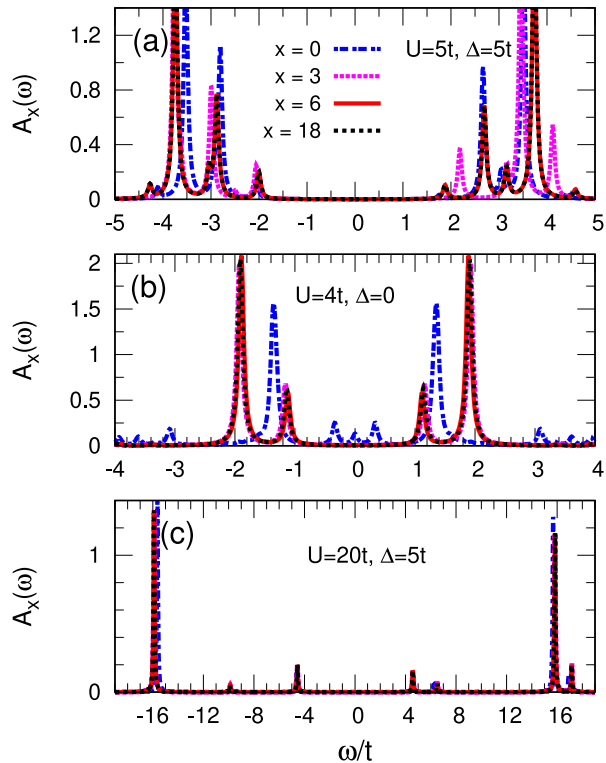}
   \caption{The spectral function $A_x(\omega)$ given by Eq.\ \eqref{eq:sf} vs frequency for a cylinder of the size
   $41\times40$ with the edges at $x=0$ and at $x=40$ in the trivial band insulator phase for $U=5t$ and $\Delta=5t$ (a),
   in the quantum spin-Hall insulator phase for $U=4t$ and $\Delta=0$ (b), and in the N\'eel antiferromagnetic phase 
   with the spin polarization in the $x$-$y$ plane for $U=20t$ and $\Delta=5t$ (c). The results are symmetric with
   respect to the middle $x=20$ of the cylinder. The results are obtained for the number of bath sites 
   $n_{bs}=6$.}
   \label{fig:sf}
   \end{center}
\end{figure}

The real-space DMFT permits to spot bulk and edge excitations of the interacting system
by considering cylindrical geometries. Although due to the finite number of bath sites in the ED impurity solver 
the spectral function is not smooth consisting of separate sharp peaks, it can still signal the existence 
of gapless excitations at the edges \cite{Hafez-Torbati2020,Ebrahimkhas2021}. This is observed also for the ED 
of lattice models \cite{Varney2010}. Fig.\ \ref{fig:sf} displays the spectral function resolved along the $x$ direction 
$A_x(\omega)$ for a cylinder of the size $41\times 40$ with the edges at $x=0$ and at $x=40$. {The results in Fig.\ \ref{fig:sf} 
correspond to} the BI phase for $U=5t$ and $\Delta=5t$ (a), to the QSHI phase for $U=4t$ and $\Delta=0$ (b),
and to the $xy$-AF phase for $U=20t$ and $\Delta=5t$ (c). 
The spectral function $A_x(\omega)$ is given by 
\be
\label{eq:sf}
A_x(\omega)=\frac{1}{4} \sum_{\sigma} \left(
A_{x,y;\sigma}(\omega)+A_{x,y+1;\sigma}(\omega)
\right) \ ,
\ee
where $A_{x,y;\sigma}(\omega)$ is the spectral function at the lattice position $(x,y)$ {for} the spin $\sigma$.
One notes that $A_x(\omega)$ does not depend on $y$ and $\sigma$ as we average over the two sites in the unit cell
in the $y$ direction and over the spin. 
We have limited the $x$ coordinate in Fig.\ \ref{fig:sf} to $x<20$ as the results are symmetric with respect to 
the middle $x=20$ of the cylinder. 
The spectral function in Eq.\ \eqref{eq:sf} averaged over the spin and the periodic 
$y$ direction but resolved along the open $x$ direction allows us to examine how the charge excitations change from the 
bulk to the edge of the system. The results are for the number of bath sites $n_{bs}=6$. A Lorentzian 
broadening with the broadening factor $0.05t$ is used in the calculations.

\begin{figure}[t]
   \begin{center}
   \includegraphics[width=0.45\textwidth,angle=0]{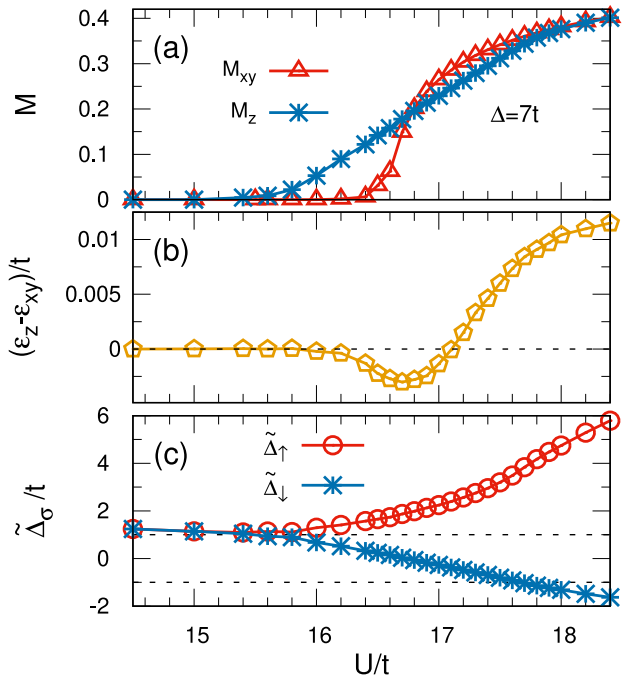}
   \caption{(a) The local spin polarization in the $xy$-AF state ($M_{xy}$) and in the $z$-AF state ($M_{z}$) 
   vs the Hubbard interaction $U$. (b) The energy difference per lattice site of the $z$-AF state and 
   the $xy$-AF state vs $U$. (c) The renormalized staggered potential $\tilde{\Delta}_\sigma$ defined 
   by Eq.\ \eqref{eq:rp} in the $z$-AF phase vs the Hubbard $U$. The dashed lines in panel (c) separate the 
   topological region $|\tilde{\Delta}_{\sigma}|<t$ from the trivial region $|\tilde{\Delta}_{\sigma}|>t$.
   The results are for the staggered potential $\Delta=7t$. We have used the number of bath sites 
   $n_{bs}=6$ in the ED impurity solver.
   The results in panel (c) are for the magnetic solution with the spin-up particles occupying
   mainly the lower-energy sublattice.}
   \label{fig:d7}
   \end{center}
\end{figure}

As one can see from Fig.\ \ref{fig:sf}(b) there are gapless excitations at the edges which quickly disappear 
as the bulk is approached. In contrast, we find gapped excitations in the bulk and at the edges in Figs. \ref{fig:sf}(a)
and \ref{fig:sf}(c). 
As we move away from the edges the results in all the panels in Fig.\ \ref{fig:sf} perfectly coincide with the 
bulk results $x=18$. 
Fig.\ \ref{fig:sf} allows us to directly investigate the topological and the trivial nature of the different 
phases in the interacting system without using the topological Hamiltonian method, if we assume that 
the bulk-boundary correspondence is valid. 
However, one notes that this approach is only accurate when the system is deep within a phase, 
where the edge excitations is either gapless as in Fig.\ \ref{fig:sf}(b) or there is a large 
energy gap as in Figs.\ \ref{fig:sf}(a) and \ref{fig:sf}(c).
Near the phase boundaries one needs a high resolution energy spectrum to distinguish a gapless spectrum 
from a spectrum which has a tiny energy gap. In our analysis, the energy resolution is limited 
by the finite number of bath sites in the ED impurity solver.
To determine the topological phase boundaries accurately one needs to use the topological Hamiltonian method 
and compute the topological invariants which change abruptly at a transition point.

\section{Antiferromagnetic Chern insulator}
\label{sec:ci}

Up to now our discussion of the phase diagram in Fig.\ \ref{fig:pd} has been limited to small and intermediate values 
of $\Delta$ with $\Delta \leq 5t$. 
In the following we will focus on large values of $\Delta$ where a $z$-AFCI appears between the BI and the $xy$-AF phase.

\begin{figure}[t]
   \begin{center}
   \includegraphics[width=0.45\textwidth,angle=0]{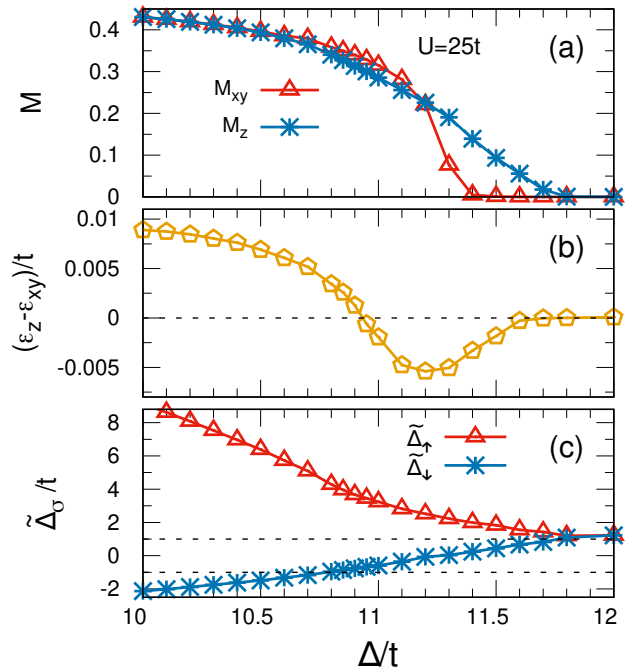}
   \caption{The same quantities as in Fig.\ \ref{fig:d7} but for the fixed Hubbard interaction $U=25t$
   and varying {staggered} potential $\Delta$.}
   \label{fig:u25}
   \end{center}
\end{figure}

In Fig.\ \ref{fig:d7}(a) the local spin polarizations in the $z$-AF and in the $xy$-AF states are 
depicted vs the Hubbard $U$ for the staggered potential $\Delta=7t$. The number of bath sites is given by $n_{bs}=6$. 
Comparing with the results in
Fig.\ \ref{fig:wd}(a) one can see how a region with a finite $M_z$ and zero $M_{xy}$ develops
upon increasing the value of $\Delta$. The energy difference between the $z$-AF state and the $xy$-AF state
plotted vs $U$ for $\Delta=7t$ in Fig.\ \ref{fig:d7}(b) shows that the $z$-AF state {becomes the stable phase
with $\varepsilon_z<\varepsilon_{xy}$} before $M_{xy}$ vanishes. 
A very similar spin-flop phase transition between the $xy$-AF order and the $z$-AF order is 
observed in the Kane-Mele-Hubbard model \cite{Jiang2018}. 
The phase diagram of the Kane-Mele-Hubbard model 
is investigated also in Ref.\ \cite{Triebl2016}. However, the analysis in Ref.\ \cite{Triebl2016} is limited 
to small values of the sublattice potential and the $z$-AF state is never found as the stable phase.

To address the topological properties of the $z$-AF phase we have plotted 
the renormalized staggered potential \eqref{eq:rp} vs the Hubbard interaction in Fig.\ \ref{fig:d7}(c)
for $\Delta=7t$. 
{The dashed lines separate the topological $|\tilde{\Delta}_{\sigma}|<t$ from the trivial $|\tilde{\Delta}_{\sigma}|>t$ region.}
In the paramagnetic regime for $U<15.5t$ the renormalized staggered potential 
$\tilde{\Delta}_\sigma$ is independent {of the} spin and $\tilde{\Delta}_\sigma >t$ suggests a BI phase.   
At $U\approx 15.5t$ we find $\tilde{\Delta}_\sigma$ extremely close to $t$. As the system 
enters the $z$-AF phase for $U>15.5t$ the renormalized staggered potential $\tilde{\Delta}_\sigma$ 
becomes spin dependent, increasing for one spin component and decreasing for the other. 
One spin component falls in the trivial region $|\tilde{\Delta}_\sigma|>t$ and the other in the topological 
region $|\tilde{\Delta}_\sigma|<t$.
The results provided in Fig.\ \ref{fig:d7}(c) correspond to the $z$-AF solution with the lower-energy 
sublattice occupied mainly with spin up particles. 
Fig.\ \ref{fig:d7}(c) demonstrates that the $z$-AF state has a finite Chern number and the emergence 
of the $z$-AFCI. 

To further check the existence of the $z$-AFCI between the BI and the $xy$-AF phase for $(U\sim 2\Delta)\gg t$ 
we have investigated in Fig.\ \ref{fig:u25} the same quantities as in Fig.\ \ref{fig:d7} but for the fixed Hubbard 
interaction $U=25t${,} varying the staggered potential $\Delta$. One can see a very similar behavior as 
in Fig.\ \ref{fig:d7}. 
The $z$-AF phase {is stable} for $10.9t\lesssim \Delta \lesssim11.8t$ with one spin component in the 
topological region $|\tilde{\Delta}_{\sigma}|<t$ and the other in the trivial region $|\tilde{\Delta}_{\sigma}|>t$.
Similar to Fig.\ \ref{fig:d7} we find that at the point where $M_z$ vanishes the renormalized staggered 
potential $\tilde{\Delta}_\sigma$ is {very} close to $t$. {However, we note that we cannot
in essence rule out the existence of an extremely narrow trivial $z$-AF phase between the BI and the 
$z$-AFCI which corresponds to a renormalized staggered potential $\tilde{\Delta}_\sigma$ being slightly 
above $t$ rather than being {\it exactly} at $t$ at the point where $M_z$ vanishes}.

We have carried out the same analysis as in Fig.\ \ref{fig:u25} also for the Hubbard interactions $U=15t$, $20t$, and $27t$ 
and the obtained transition points are specified in the phase diagram Fig.\ \ref{fig:pd}.
Fig.\ \ref{fig:pd} exhibits a constant width proportional to $t$ for the $z$-AFCI as 
the atomic limit $U/t,\Delta/t \to \infty$ is approached.
In the phase diagram of the Kane-Mele-Hubbard model obtained using mean-field theory, the width 
of the $z$-AFCI phase increases as the Hubbard $U$ and the sublattice potential $\Delta$ are increased \cite{Jiang2018}.
In the slave boson mean-field analysis of the Kane-Mele-Hubbard model, however, a constant width for the $z$-AFCI {phase} is 
found \cite{Jiang2018} quite comparable to our results in Fig.\ \ref{fig:pd}. This 
close similarity of the results obtained for these two different models shows that the AFCI is a robust phase 
which can emerge between the BI and the trivial AF Mott insulator independent from the details 
of the system. What is essential is the presence of the spin-orbit coupling and the absence of a space group operation that could
compensate the effect of the time-reversal transformation on the electronic state. 

\begin{figure}[t]
   \begin{center}
   \includegraphics[width=0.42\textwidth,angle=0]{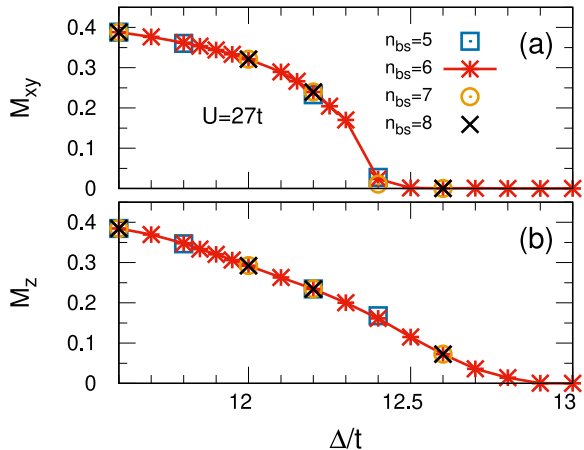}
   \caption{Comparison of the local spin polarizations obtained for different number of bath sites $n_{bs}$ in the 
   ED impurity solver for the Hubbard interaction $U=27t$ and varying staggered potential $\Delta$ across the $z$-AFCI phase.
   Panels (a) and (b) display the local spin polarizations in the $xy$-AF state ($M_{xy}$) and 
   in the $z$-AF state ($M_{z}$).}
   \label{fig:bs}
   \end{center}
\end{figure}

To show that the number of bath sites in the impurity solver does not alter our results 
we have compared in Fig.\ \ref{fig:bs} the local spin polarizations $M_{xy}$ (a) and $M_z$ (b) 
for $n_{bs}=6$ at selective points across the $z$-AFCI {phase} with the results obtained for $n_{bs}=5$, 
$7$, and $8$. The results are for the Hubbard interaction $U=27t$. One can hardly see any change 
in the data upon changing the number of bath sites. 

\section{Antiferromagnetic Chern insulator from gap closing perspective}
\label{sec:gap}

\begin{figure}[t]
   \begin{center}
   \includegraphics[width=0.48\textwidth,angle=0]{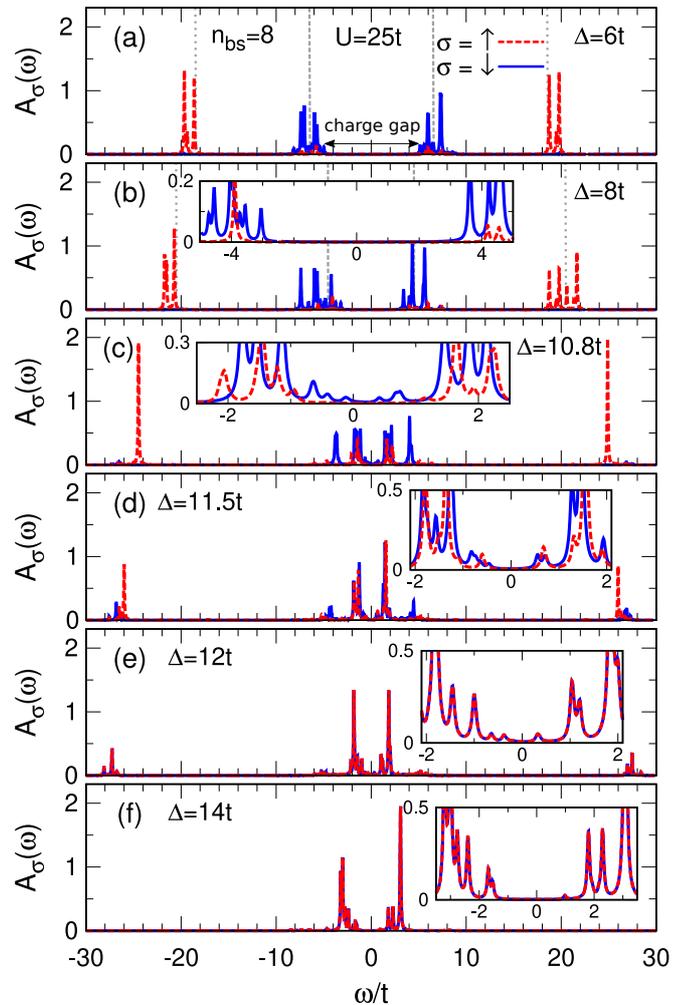}
   \caption{The spin-resolved spectral function $A_\sigma(\omega)$ given by Eq.\ \eqref{eq:sfb} vs frequency 
   for the Hubbard interaction $U=25t$ and different values of the staggered potential $\Delta$ across the 
   $z$-AFCI phase. The results in panels (a) to (d) correspond to the $z$-AF solution with the lower energy 
   sublattice mainly occupied with spin up particles, see panel (a) in Fig.\ \ref{fig:cg}. The results are for 
   the number of bath sites $n_{bs}=8$ in the ED impurity solver.}
   \label{fig:sfb}
   \end{center}
\end{figure}

\begin{figure}[t]
   \begin{center}
   \includegraphics[width=0.45\textwidth,angle=0]{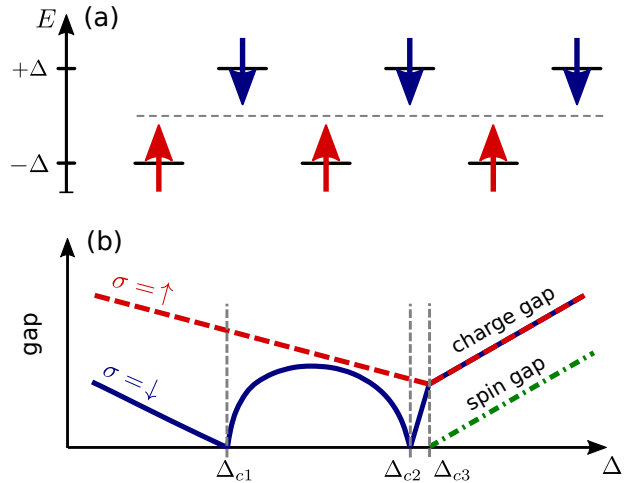}
   \caption{(a) Schematic representation of an AF state with the spin-up sublattice having a lower on site energy
   ($E=-\Delta$) in contrast to the spin-down sublattice ($E=+\Delta$). 
   (b) Schematic representation of the charge gap for the spin up ($\sigma=\uparrow$) and the 
   spin down ($\sigma=\downarrow$) vs the staggered potential $\Delta$ across the $z$-AFCI. 
   The sketch corresponds to the $z$-AF solution in which the sublattice with the lower onsite energy 
   is mainly occupied with spin up particles, as shown in panel (a). 
   The spin gap in the trivial band insulator phase, corresponding to $\Delta>\Delta_{c3}$, is also shown in panel (b). 
   {The spin gap stems from an electron-hole bound state with the total spin $S=1$ and the magnetic polarization 
   $\alpha=z$, see the main text}. 
   The local spin polarization $M_z$ becomes finite for $\Delta<\Delta_{c3}$
   upon closing the spin gap.
   The Chern number $\mathcal{C}_{\sigma}$ changes upon closing the charge gap for the spin component $\sigma$. 
   The Chern number is finite only for $\Delta_{c1}<\Delta<\Delta_{c2}$.}
   \label{fig:cg}
   \end{center}
\end{figure}

As the staggered potential $\Delta$ is increased in Fig.\ \ref{fig:u25}(c) the spin up component 
remains in the trivial 
region while the spin down component undergoes a transition from the trivial to the topological 
region. Such a transition requires closing of the charge gap. Hence, the charge excitations for 
the different spin components are expected to show different behaviors.
To see how the charge excitations in the system depend on the spin we plot the spin-resolved 
spectral function  
\be 
\label{eq:sfb}
A_\sigma(\omega)=\frac{1}{4} 
\sum_{(x,y)\in {\rm unit~cell}} A_{x,y;\sigma}(\omega) 
\ee
in Fig.\ \ref{fig:sfb}. The sum in Eq.\ \eqref{eq:sfb} runs over the $2\times 2$ unit cell specified 
in Fig.\ \ref{fig:hhh}. The results in Fig.\ \ref{fig:sfb} are given for $U=25t$ and the number of 
bath sites $n_{bs}=8$. Similar to the results in Fig.\ \ref{fig:u25}(c), the results in Fig.\ \ref{fig:sfb} are 
for the magnetic solution with the lower-energy sublattice mainly occupied with spin up particles, see
Fig.\ \ref{fig:cg}(a) for a schematic sketch of the state. 
The magnetic results in Fig.\ \ref{fig:sfb} are always given for the $z$-AF solution, 
even for the small values of $\Delta=6t$ in Fig.\ \ref{fig:sfb}(a) and $\Delta=8t$ in Fig.\ \ref{fig:sfb}(b) 
where the $z$-AF state is metastable, see Fig.\ \ref{fig:u25}(b). We avoid the spin-flop transition 
because our aim in this section is to study the stabilization of the $z$-AFCI phase from the perspective 
of gap closing, i.e., to study how the charge gap continuously changes as $\Delta$ is increased in 
Fig.\ \ref{fig:u25}(c) and the transition to the $z$-AFCI phase takes place.

The staggered potentials $\Delta=6t$ and $8t$ in panels (a) and (b) in Fig.\ \ref{fig:sfb} correspond 
to a highly polarized $z$-AF phase with both spin components in the trivial region. For the fully polarized 
state the charge excitation for the spin up corresponds to moving a particle from the lower-energy 
sublattice to the higher-energy sublattice, see Fig.\ \ref{fig:cg}(a). This costs an energy equal to $U+2\Delta$. 
This nicely explains the spectral weights  for spin up
in panels (a) and (b) in Fig.\ \ref{fig:sfb} distributed mainly near $\omega=\pm(U/2+\Delta)$.
The points $\omega=\pm(U/2+\Delta)$ in Figs.\ \ref{fig:sfb}(a)
and \ref{fig:sfb}(b) are specified by vertical dotted lines.
The charge excitation for spin down corresponds to moving a particle from the higher-energy sublattice 
to the lower-energy sublattice, which costs an energy equal to $U-2\Delta$. This explains the 
spectral weight distribution for spin down appearing around $\omega=\pm(U/2-\Delta)$ in panels (a) and (b) in 
Fig.\ \ref{fig:sfb}. 
The points $\omega=\pm(U/2-\Delta)$ in Figs.\ \ref{fig:sfb}(a) and \ref{fig:sfb}(b) are specified by 
vertical dashed lines.
One notes that near $\omega=\pm(U/2-\Delta)$ there are also some spectral weights 
for spin up which can more obviously be seen from the inset in panel (b). These contributions are important
because they define the charge gap for the spin up and originate from the fact that the system is not fully polarized.

The charge gap is defined by the energy difference of the electron and the hole spectral peaks closest 
to the Fermi energy $\omega=0$. The charge gap is specified by a double-head arrow in Fig.\ \ref{fig:sfb}(a). 
Panels (a) and (b) in Fig.\ \ref{fig:sfb} show that the charge gap for the spin down is always smaller 
than the charge gap for the spin up. The charge gap for both spin components decreases as the staggered 
potential $\Delta$ is increased approaching the transition point to the $z$-AFCI.
According to Fig.\ \ref{fig:u25}(c) the spin down component undergoes a transition from the trivial state to the 
topological state at $\Delta=10.8t$. Panel (c) in Fig.\ \ref{fig:sfb}
displays the spin-resolved spectral function for the same value of $\Delta$. As one can see from the inset 
in panel (c) there are some spectral contributions near the Fermi energy for the spin down component while 
the spin up component is clearly gapped. This is nicely consistent with the expectation that the change 
of the Chern number is accompanied by closing of the charge gap.

For the staggered potential $\Delta=11.5t$ in Fig.\ \ref{fig:sfb}(d) the system is within 
the $z$-AFCI. The spectral weight distribution for the spin up shows a 
manifest shift of the spectral weights from $\omega \approx \pm(U/2+\Delta)$ to $\omega \approx \pm(U/2-\Delta)$ 
in contrast to the spectral weight distribution for small values of $\Delta$. 
This is due to the noticable reduction of the local spin polarization $M_z$ and approaching the paramagnetic region
as can be seen from Fig.\ \ref{fig:u25}(a). The inset in Fig.\ \ref{fig:sfb}(d) suggests a finite and rather 
equal charge gap for both spin components. This finite charge gap corroborates the insulating bulk state for 
both spin components. For the staggered potential $\Delta=12t$ in Fig.\ \ref{fig:sfb}(e) the system enters 
the paramagnetic phase and the spectral functions for up and down spins coincide. We observe an increase in 
the charge gap upon increasing the staggered potential from $\Delta=12t$ in Fig.\ \ref{fig:sfb}(e) to $\Delta=14t$ 
in Fig.\ \ref{fig:sfb}(f) which signals a BI phase consistent with our finding based on the renormalized staggered 
potential in Fig.\ \ref{fig:u25}(c).

Although it is not possible to extract from the ED spectral functions quantitative values for the charge gap which changes smoothly across the narrow 
$z$-AFCI region, a qualitative behavior can still be concluded. 
One should note that apart from the finite number of bath sites in our ED impurity solver, it is 
inherently difficult to obtain accurate results especially for dynamical quantities near a critical region.
Fig.\ \ref{fig:cg}(b) shows schematically how the different phase transitions take place 
upon changing the staggered potential across the $z$-AFCI from the gap closing perspective. In addition to 
the charge gap for the up and the down spin the figure includes also the spin gap in the BI phase. The spin gap 
is the excitation energy of an electron-hole bound state with the total spin $S=1$.
It is the condensation of such a bound state which leads to the stabilization of an AF order. 
The stabilization of an AF state upon
softening an $S=1$ excitation is well-known in dimerized magnetic systems
\cite{Sachdev1990,Rueegg2003,Rueegg2004,Nikuni2000a,Fischer2011}. 
It can also be seen from the 
analysis of verious gaps for the ionic Hubbard model \cite{Hafez-Torbati2016} as an electronic model similar to
our Hamiltonian in Eq.\ \eqref{eq:hhh}. One should note, however, that there is a spin-dependent hopping phase in Eq.\ \eqref{eq:hhh} 
which lifts the degeneracy between the magnetic 
number $m=0$ and the magnetic numbers $m=\pm1$, or equivalently in terms of a specific magnetic polarization $\alpha=x,y,z$ 
\cite{Sachdev1990}, 
the degeneracy between the magnetic polarization $\alpha=z$ and the magnetic polarizations $\alpha=x$,$y$.
It is the magnetic polarization $\alpha=z$ which defines the spin gap in Fig.\ \ref{fig:cg}(b) and stabilizes the 
$z$-AF state \cite{Sachdev1990}. 
In sketching the charge gap for the up and the down spins in Fig.\ \ref{fig:cg}(b) 
it is supposed that the spin-up sublattice has the lower onsite energy in contrast to the spin-down sublattice, 
see Fig.\ \ref{fig:cg}(a).

Starting from the limit of large values of $\Delta$ in Fig.\ \ref{fig:cg}(b) the system is in the BI phase 
with an identical charge gap for the up and the down spins. Upon decreasing $\Delta$ the spin gap vanishes at $\Delta_{c3}$ 
and the system acquires a $z$-AF order for $\Delta<\Delta_{c3}$. 
The charge gap becomes spin dependent in the magnetically ordered phase. 
This is due to the fact that the effect of the time-reversal transformation on the $z$-AF state cannot be
compensated by a space group operation. Otherwise, the charge gap would remain 
equal for the up and the down spins even in a magnetically ordered phase. 
This would prevent the different spin components to fall in different topological states.

The charge gap for the spin down in Fig.\ \ref{fig:cg}(b) closes at $\Delta_{c2}$ and the system 
acquires a finite Chern number. This leads to the $z$-AFCI. 
Upon further decreasing of $\Delta$ the charge gap for the spin down closes again at $\Delta_{c1}$. 
The Chern number of the spin down component vanishes for $\Delta<\Delta_{c1}$ and the system enters 
a trivial $z$-AF phase. 

Our description of the phase transitions in Fig.\ \ref{fig:cg}(b) implies 
a trivial $z$-AF phase between the BI and the $z$-AFCI, 
although in practice it might be extremely narrow. 
This is because the magnetic ordering stems from the condensation of an $S=1$ bound state while the 
change of the Chern number requires closing of the charge gap. 
This corresponds to a renormalized staggered potential in Fig.\ \ref{fig:u25}(c) being slightly 
above $t$ rather than being exactly at $t$ at the point where $M_z$ vanishes 
in Fig.\ \ref{fig:u25}(a) and cannot be resolved based on numerical DMFT results. 
An intervening trivial AF phase between the BI and the AFCI in the case of a continuous transition is 
also pointed out in Ref.\ \cite{Jiang2018} based on bifurcation of a Weyl line due to the AF ordering.
Further research is needed to clarify the emergence of the AFCI based on vanishing 
of different energy gaps proposed in Fig.\ \ref{fig:cg}(b). 
Using the numerical renormalization group instead of the ED as the impurity solver in DMFT 
might help to better resolve the behavior of the charge gap across the AFCI.
Beyond the DMFT, a method such as the continuous unitary transformations  
\cite{Knetter2000,Krull2012,Powalski2015} 
which provides a quantitative description of the elementry excitations and their interactions can 
shine more light on verious excitation energies, including both the charge and the spin gap, 
and the stabilization of the AFCI.
The method is already applied to the ionic Hubbard model \cite{Hafez-Torbati2016} and can be employed 
to address also the effect of the finite spin-orbit coupling.

\section{conclusion}
\label{sec:con}
In this paper we demonstrate the existence of the collinear AFCI in a square lattice system which 
preserves the inversion symmetry. 
Our analysis relies on a minimal extension of the time-reversal-invariant Harper-Hofstadter-Hubbard (TRI HHH) model 
which permits to study the competition of the BI, the QSHI, and the N\'eel AF Mott insulator phases by varying the 
Hubbard repulsion $U$ and a sublattice potential $\Delta$ and examine the existence of the AFCI phase 
beyond the noncentrosymmetric Kane-Mele-Hubbard model \cite{Jiang2018}.

We map out the phase diagram of the model in the $U$-$\Delta$ plane showing a very close similarity with 
the phase diagram proposed for the noncentrosymmetric Kane-Mele-Hubbard model \cite{Jiang2018}. 
The AFCI appears between the BI and the AF Mott insulator in the limit of large $U$ and $\Delta$. 
The close similarity of the results obtained for the two very different models suggests that the AFCI 
is a generic phase which exists beyond a specific model or a lattice structure.
We discuss the stabilization of the AFCI based on the vanishing of the charge and the spin gap 
and emphasize that the 
necessary condition to realize the AFCI is the lack of a space group operation to compensate the 
effect of the time-reversal transformation on the electronic state, which can be achieved on different 
lattice structures with or without the inversion symmetry.

Optical lattices provide a versatile platform to study the interplay of two-particle 
interaction and spin-orbit coupling \cite{Hofstetter2018}.
The correlation strength is highly tunable by using Feshbach resonances or by changing the optical lattice depth, 
and there has been impressive progress in the past decade in creation of artificial gauge fields 
using, for example, lattice shaking or laster-assisted tunneling techniques \cite{Aidelsburger2018}. 
Fundamental topological models such as the Haldane model \cite{Jotzu2014} and the TRI Harper-Hofstadter 
model \cite{Aidelsburger2013,Miyake2013} 
have thus been realized. The Chern number of the Hofstadter bands has been measured \cite{Aidelsburger2014} and the 
phase diagram of the Haldane model has been mapped out \cite{Jotzu2014}. 
The high control and tunability of quantum gases in optical lattices would allow to directly investigate 
our proposed phase diagram in Fig.\ \ref{fig:pd}. According to our results, the AFCI can be realized by driving 
the system from the AF Mott insulator to the BI phase in the limit of large $U$ and $\Delta$.

Since the experimental discovery of the quantum anomalous Hall effect in Cr-doped (Bi,Sb)$_2$Te$_3$ thin films
about a decade ago \cite{Chang2013} there has been extensive research to increase the observation temperature
of the effect \cite{Wang2021,Liu2021}.
In the current realizations of the CI the magnetic ordering and the electronic properties  
stem from electrons in different orbitals \cite{Liu2016,Tokura2019,Wang2021,Liu2021}. While the strongly 
interacting electrons in a $3d$ or $4f$ orbital are the origin of the magnetic ordering, the non-interacting 
electrons in $6p$ and $5p$ orbitals define the electronic properties \cite{Li2019b}. 
The coupling between the magnetic and the electronic degrees of freedom can be seen as an AF Kondo interaction \cite{Yoshida2016}.
In contrast, in the AFCI phase we discussed in this paper the electronic and the magnetic properties 
are inherently coupled as they originate from the same, strongly interacting, electrons. 
Such an strongly correlated AFCI is expected to show a strong magnetic blue shift of the charge gap upon developing the 
magnetic order below the N\'eel temperature \cite{Bossini2020,Hafez-Torbati2021}.     
This would allow to realize the quantum anomalous Hall effect at temperatures closer to the magnetic 
transition temperature of the material. 
Transition metal elements with partially filled $4d$ and $5d$ shells such as Iridates \cite{Rau2016,Bertinshaw2019} 
are potential candidates to 
observe the combined effect of the strong correlation and the spin-orbit coupling and to realize 
the AFCI phase.

\acknowledgements
We would like to thank Y. Xu, A. Dutta, and I. Titvinidze for useful discussions.
This work  was  supported  by  the  Deutsche  Forschungsgemeinschaft 
(DFG,  German  Research  Foundation)  
via  Research  Unit  FOR  2414  under  Project No.  277974659 (M.E. and W.H.). 
This  work was also supported financially in TRR 160 (M.H.-T. and G.S.U.) by the DFG and 
via the high-performance computing center 
Center for Scientific Computing (CSC).

\section*{references}

%

\end{document}